\documentclass{article}
\usepackage{spconf,amsmath,graphicx}
\usepackage{bm, amssymb, booktabs, multirow,cite, url, xcolor, hyperref}
\usepackage{makecell, arydshln}

\title{DiaCorrect: ERROR CORRECTION BACK-END FOR SPEAKER DIARIZATION}

\name{\begin{tabular}{c}Jiangyu Han$^{1}$, Federico Landini$^{1}$, Johan Rohdin$^{1}$, Mireia Diez$^{1}$, Luk\'{a}\v{s} Burget$^{1}$ \\ Yuhang Cao$^{2}$, Heng Lu$^{2}$, Jan \v{C}ernock\'{y}$^{1}$\end{tabular}}
\address{$^1$Brno University of Technology, Faculty of Information Technology, Speech@FIT, Czechia\\
$^2$Ximalaya Inc., Shanghai, China}

\begin{document}
\ninept
\maketitle

\begin{abstract}
In this work, we propose an error correction framework, named DiaCorrect, to refine the output of a diarization system
in a simple yet effective way. 
This method is inspired by error correction techniques in automatic speech recognition.
Our model consists of two parallel convolutional encoders and a transformer-based decoder. 
By exploiting the interactions between the input recording and the initial system's outputs, DiaCorrect can automatically
correct the initial
speaker activities to minimize the diarization errors.  
Experiments on 2-speaker telephony data 
 show that the proposed DiaCorrect can
effectively improve the initial model's results. 
Our source code
is publicly available at \url{https://github.com/BUTSpeechFIT/diacorrect}.

\end{abstract}
%

\section{Introduction}
\label{sec:intro}
\vspace{-0.1cm}




Speaker diarization aims to address ``who spoke when" by determining speaker-specific
time-segments in a long multi-speaker recording \cite{park2022review}.
Traditionally, it is performed by 
clustering short speech segments according to their speaker identity \cite{sell2014speaker, garcia2017speaker, landini2022bayesian}. In recent years, end-to-end neural diarization (EEND)\cite{fujita19_interspeech} and its extensions \cite{fujita2019end, horiguchi20_interspeech, fujita2023neural} have shown excellent performance when handling overlapped speech. 
However, end-to-end models require large amounts of training data. Given the time-consuming and expensive nature of producing manual labels, the compromise solution consists in generating training data artificially~\cite{fujita19_interspeech,yamashita22_odyssey,landini2022simulated, landini2023multi}. Due to the existent mismatch between artificial and real target data, models trained on artificial data are frequently fine-tuned on a real development set to achieve good results. Nevertheless, ground-truth annotations for target data are not always available and, even more, if the practitioner does not have access to the model but only to its outputs, fine-tuning is impossible.  In such scenarios, an error correction model can automatically refine and correct the initial diarization results.

In automatic speech recognition (ASR), error correction is a typical sequence-to-sequence task. It takes a sentence generated by an ASR model, and aims to correct the errors in the source
sequence to match the ground-truth transcriptions\cite{errattahi2018automatic, leng2021fastcorrect}. Error correction techniques have been widely adopted to refine the initial ASR results for further word error rate reduction. 
In \cite{guo2019spelling}, authors first proposed an LSTM-based correction model by training on synthesized speech generated from a text-only corpus. 
Since then, 
several related works have been proposed,
such as a
transformer corrector \cite{zhang2019investigation} for Mandarin speech recognition, 
using acoustic features and the recognized results for code-switching ASR \cite{zhang2021end}, and initialization from pre-trained models \cite{hrinchuk2020correction} for better correction performance.


Despite the error correction techniques being common in ASR, to the best of our knowledge, 
there are only a few related studies in speaker diarization.
In \cite{horiguchi2021end},  
a post-processing method with an EEND model for clustering-based diarization is proposed.
The method iteratively selects frames from all speaker pairs and processes them with a pre-trained EEND model to find overlapped speech regions that the clustering-based system could not handle.
More recently, authors in \cite{paturi2023lexical} propose to correct word-level speaker error labels using lexical information. 
A pre-trained language model is required to
infuse the lexical knowledge to correct speaker errors while leveraging speaker scores from the diarization system to prevent over-corrections. The major limitation of this method is its strong coupling with ASR tasks. 



In this paper, we explore, whether it is possible to refine diarization results similar to the error correction in ASR: as long as an initial result is provided, corrections can be conducted without excessive constraints. 
Motivated by error correction in ASR, we propose an error correction framework for speaker diarization which we name DiaCorrect.
It automatically refines the diarization results provided by an initial diarization system, by 
exploiting the interactions between the input acoustic features 
and the
initial speaker activity predictions (SAPs) with two parallel convolutional encoders and 
a transformer-based decoder.
In this work, we use EEND-EDA~\cite{horiguchi20_interspeech} to provide initial diarization outputs but DiaCorrect can be applied on top of any diarization systems.
To reduce over-fitting in the error correction process, we adopt data pruning to select hard samples from the simulated training set. Moreover, by analyzing the distribution of initial SAPs, we calibrate them for the inference.
We analyze DiaCorrect's performance under two scenarios. If only simulated data are available, DiaCorrect can be trained on a small set of hard samples. 
If some limited target domain samples are provided, DiaCorrect can be fine-tuned or even trained from scratch on such data.
Experiments on 2-speaker telephony data, CALLHOME~\cite{callhome} and DIHARD-\uppercase\expandafter{\romannumeral3}~\cite{ryant21_interspeech} CTS, indicate that our method is effective to improve the initial diarization performance in real scenarios, even when only using less than 30\,hours of simulated training data.

\section{Methods}
\label{sec:eend}
\vspace{-0.1cm}
\subsection{EEND baseline}
\vspace{-0.1cm}

We take the self-attention EEND
with encoder-decoder attractors (EEND-EDA) \cite{horiguchi20_interspeech} as our baseline because of its superior performance in previous works. 
Different from clustering-based methods, EEND reformulates speaker diarization as a multi-label classification task, and directly
outputs the joint speech activities of all speakers at the same time for an input
multi-speaker recording. Permutation-invariant training (PIT) \cite{kolbaek2017multitalker} is applied to solve the output label ambiguity problem. More details can be found in \cite{horiguchi20_interspeech}.


\subsection{Proposed DiaCorrect}
\label{ssec: diacorrect}
\vspace{-0.1cm}
\subsubsection{Motivation}
\vspace{-0.1cm}

Motivated by error correction techniques in ASR, the proposed DiaCorrect aims to improve the diarization results from the initial model
outputs. Although the contextual patterns are not as prominent as in ASR, 
we still believe that it is possible to 
learn some hidden patterns across frames.
For example, the model could learn global statistics for the whole conversation and have a notion of how likely each speaker is to speak next.
Besides, it should be unlikely for very high and very low probability values to alternate frame after frame.

In addition, we believe that the additional audio features are able to provide rich acoustic information for refining the initial speaker activities. Therefore, DiaCorrect processes both acoustic features and initial speaker activities to produce the corrected predictions.

\subsubsection{Math definition}
\vspace{-0.1cm}
Given a $T$-length, $F$-dimensional input acoustic feature sequence $X = (\bm{x_t} \in \mathbb{R}^F \ |\ t = 1, ..., T)$ and its initial SAPs $Z = (\bm{z_t} \in \mathbb{R}^C|\ t = 1, ..., T)$ of $C$ speakers produced by an external diarization system, DiaCorrect aims to map the initial SAPs to the labels
${Y = (\bm{y_t} \in \{0, 1\}^C \ |\ t = 1, ..., T)}$.

During DiaCorrect error correction, the refined speaker activity predictions $\tilde{Z}$ are generated as follows\footnote{Note that we follow similar notation to EEND works~\cite{fujita2019end,horiguchi20_interspeech} and the main change is the addition of $Z$ as another input.}:
\begin{equation}
\label{eq:1}
\tilde{Z} = \underset{\hat{Z} \ \in \ \mathcal{Z}}{\arg \max} \ P(\hat{Z} | X, Z)
\end{equation}
where $\hat{Z} = (\bm{\hat{z}_t} \in [0, 1]^C |\ t = 1, ..., T)$ is the output speaker activity prediction, $\mathcal{Z}$ is the set of all possible output predictions.
With the approximate conditional independence assumption, $P(\hat{Z}|X,Z)$ can be factorized as:
\begin{equation}
\label{eq:2}
\begin{split}
P(\hat{Z}|X, Z)&= \prod_t P(\bm {\hat{z}_t} | \bm{\hat{z}_1}, ..., \bm{\hat{z}_{t-1}}, X, Z) \\
 &\approx \prod_t P(\bm {\hat{z}_t} | X, Z) \approx \prod_t \prod_c P({\hat{z}_{t,c}} | X, Z)
\end{split}
\end{equation}


During the training stage, we also use PIT \cite{kolbaek2017multitalker} to handle the output label ambiguity.  With a standard binary cross entropy (BCE), the whole DiaCorrect loss function is defined as follows:
\begin{equation}
\label{eq:loss}
\mathcal{L} = \frac{1}{TC} \mathop{\min}_{\phi \in \mathcal{P}} \sum_{t=1}^{T}{\rm BCE}(\bm{\hat z_t^{ \phi}}, \bm{y_t})
\end{equation}
where $\mathcal{P}$ is the permutation set of $C$ speakers, $\phi$ is one of the permutations, $\bm{\hat z_t^{\phi}}$ means the output speaker activity prediction with permutation $\phi$ at time $t$.

\begin{figure}[t]
  \centering
  \includegraphics[width=8.6cm]{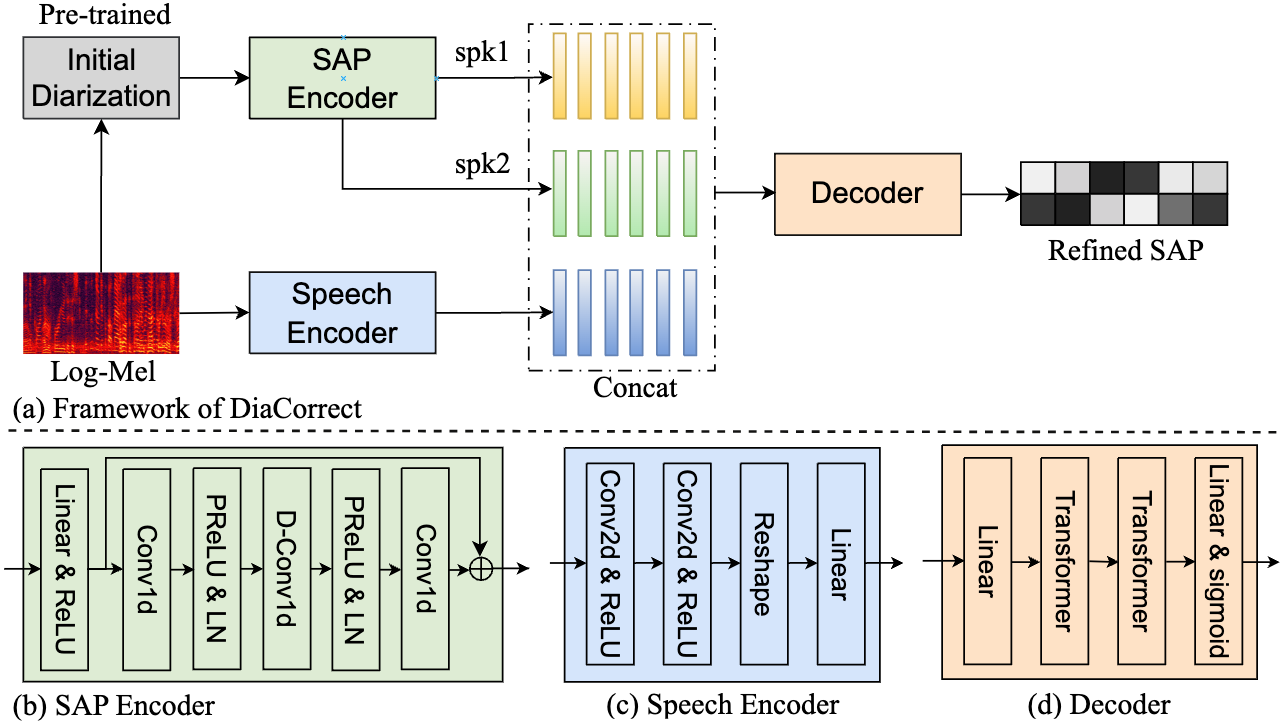}
  \caption{Framework of DiaCorrect for 2-speaker scenario.}
  \label{fig:diacorrect}
  \vspace{-0.3cm}
\end{figure}

\subsubsection{Model design}
\vspace{-0.1cm}

The architecture of DiaCorrect for 2-speaker input conversations is shown in Figure \ref{fig:diacorrect}.
It has a speaker activity prediction (SAP) encoder to leverage the speaker activity information from
the initial diarization results, and a speech encoder to capture the 
acoustic speaker interactions. The SAP encoder is designed to accept 
the logit outputs from an initial diarization system, a pre-trained EEND-EDA in this work, and process them independently for each speaker.
The input to the speech encoder are standard log-Mel features.
After transforming the initial speaker predictions and acoustic features 
into a high-dimensional embedding space, we concatenate these high-level features per frame and introduce a sequential decoder to generate refined diarization outputs based on the concatenated output. 

The structure of SAP encoder is motivated by the temporal convolutional network used in Conv-TasNet \cite{luo2019conv} to model speaker activities along time for each speaker. Figure \ref{fig:diacorrect}(b) shows the detailed structure. It mainly consists
of a 1-D convolution followed by a 1-D depthwise convolution (D-Conv1d) \cite{chollet2017xception}, with the parametric
rectified linear unit (PReLU) \cite{he2015delving} and layer normalization (LN) \cite{ba2016layer} in between them.
Skip connection \cite{he2016deep} is applied as shown in the diagram.
The input/output dimensions of the Linear layer in SAP encoder are 2/256.
The following Conv1d layer is a point-wise convolution with input/output channels of 256/512.
For the depthwise convolution layer, we set its stride/kernel size to 1/3, and input/output channels to 512/512. Then, another point-wise convolution with input/output channels of 512/256 is used.
For the speech encoder, as shown in Figure \ref{fig:diacorrect}(c), we choose 2-D convolution to model both temporal and spectral dynamics. 
The stride, kernel size, and paddings of (time, frequency) in the Conv2d layers are set to (3, 7), (1, 5), and (1, 0), respectively. The input/output dimensions of the two Conv2d and Linear layers are 1/256, 256/256, and 3328/256, respectively.
Then the convolutional outputs are reshaped to concatenate across channel and frequency dimensions. Finally, a 2-layer transformer \cite{vaswani2017attention} based decoder shown in Figure \ref{fig:diacorrect}(d) is used to predict the refined speaker activity results. 
The input/output dimensions of the two Linear layers in the decoder are set to 768/256 and 256/2, respectively.
\vspace{-0.1cm}
\section{Experiments}
\label{sec:exp}
\vspace{-0.1cm}

\subsection{Datasets}
\vspace{-0.1cm}

The EEND-EDA baseline and DiaCorrect models are trained on 2-speaker simulated conversations (SC)  \cite{landini2022simulated} that are generated using real distributions of pauses and overlaps from DIHARD-\uppercase\expandafter{\romannumeral3} \cite{ryant21_interspeech} CTS development set. 
We use 2-speaker subsets of CALLHOME\cite{callhome} Part1 (CH Part 1) and DIHARD-\uppercase\expandafter{\romannumeral3} CTS development (DH3 dev) for model fine-tuning. The 2-speaker subsets of CALLHOME Part 2 (CH Part 2) and DIHARD-\uppercase\expandafter{\romannumeral3} CTS evaluation (DH3 eval) are used to evaluate system performance in real scenarios. 
All recordings are sampled at 8 kHz.
Table \ref{tab:dataset} presents all detailed information about the datasets used in this paper.

\begin{table}[!tbp]
\vspace{-0.5cm}
  \caption{Information of 2-speaker simulated and real datasets.}
  \label{tab:dataset}
 \setlength{\tabcolsep}{1.2mm}
  \centering
  \begin{tabular}{l c c c c c c}
    \hline
        \multirow{2}{*}{Dataset} & \multirow{2}{*}{\#Utts} & \multirow{2}{*}{Hours} & Average  &  \multicolumn{3}{c}{Average \%} \\ 
        & & & dur. (s) & sil. & 1spk & over.\\
    \hline
    SC & 25k & 2480 & 356.15 & 12.80 & 78.83 & 8.37 \\
    \hline
    CH Part1 & 155 & 3.19 & 74.02 & 9.05 & 77.90 & 13.05 \\
    CH Part2 & 148 & 2.97 & 72.14 & 9.84 & 78.34 & 11.82 \\
    DH3 dev & 61 & 10.17 & 599.95 & 10.56 & 77.27 & 12.17 \\
    DH3 eval & 61 & 10.17 & 599.95 & 10.89 & 78.62 & 10.49 \\
    \hline
  \end{tabular}
  \vspace{-0.1cm}
\end{table}

\subsection{Configurations}
\vspace{-0.1cm}

We stack 
15 consecutive 23-dimensional log-scaled Mel-filterbanks (computed over 25 ms every 10 ms) to
produce 345-dimensional features.
To alleviate over-fitting to the simulated data, all DiaCorrect models are trained only for 5 epochs with Adam \cite{kingma2014adam} optimizer and an initial learning rate of 1e-5.
We average model parameters of the 5 epochs for evaluation and use a decision threshold of 0.5 for speech activities.
We report the standard diarization error rate (DER) \cite{fiscus2006rich} and its components, missed speech (Miss), false alarm (FA), and speaker confusion (Conf.).
For CH Part2 evaluation, we apply a 11-frame median filter on the outputs and we set the DER collar to 0.25\,s. For DH3, no median filter is applied and all results are evaluated with a 0\,s collar.
More details about DiaCorrect implementation can be found in our source code\footnote{\url{https://github.com/BUTSpeechFIT/diacorrect}}. 

\subsection{Results and discussion}
\vspace{-0.1cm}
\subsubsection{Baseline}
\vspace{-0.1cm}

We train the EEND-EDA\footnote{\url{https://github.com/BUTSpeechFIT/EEND}} baseline with the SC\footnote{\url{https://github.com/BUTSpeechFIT/EEND_dataprep}} dataset following the setup described in \cite{horiguchi20_interspeech, landini2022simulated}.
Results are shown in Table \ref{tab:baseline}, where ``FT'' denotes fine-tuning with the corresponding set (Part1 for CH and dev for DH3).
We first consider scenarios where target-domain data for fine-tuning is unavailable and therefore
utilize the pre-trained EEND-EDA model without fine-tuning as baseline.

\vspace{-0.3cm}
\begin{table}[!htbp]
  \setlength\tabcolsep{4pt}
  \caption{Baseline performance of EEND-EDA on real datasets.}
  \label{tab:baseline}
  \centering
  \begin{tabular}{c|cccc|cccc}
    \hline
        \multirow{2}{*}{FT} & \multicolumn{4}{c|}{CH Part2} & \multicolumn{4}{c}{DH3 eval} \\
        & DER & Miss & FA & Conf. & DER & Miss & FA & Conf. \\
    \hline
    - & 8.62 & 3.40 & 4.53 & 0.69 & 19.58 & 3.94 & 14.62 & 1.02 \\
    \checkmark & 7.88 & 5.02 & 2.18 & 0.68 & 12.76 & 8.03 & 3.88 & 0.85 \\
    \hline
  \end{tabular}
   \vspace{-0.3cm}
\end{table}

\subsubsection{DiaCorrect with data pruning}
\vspace{-0.1cm}

To train DiaCorrect, we utilize the same training data used for the baseline.
As the baseline is also the diarization system that produces the initial labels for DiaCorrect, it makes only few errors on the training data and produces speaker activity predictions that are close to the ground truth labels for most recordings.
Then, it becomes difficult to update DiaCorrect's parameters effectively and the training process is prone to overfit to the simulated conversations.

To improve correction results, we propose to select the hard recordings from the training data based on its baseline performance. 
The underlying motivation is that we believe an appropriate data pruning could reduce the model's dependence on the simulated training set, and emphasize error learning and correction during the training stage.
In this paper, the average DER of the baseline system in the training set is 3.13\%. 
To select hard samples, we choose recordings with DER between a lower and an upper limit. The upper limit is always set to 40\% to remove outliers, and the lower limit is set to 8\%, 10\%, 12\%, and 14\% to produce different subsets. The selected utterances/percentages of simulated conversations are 1254/5.0\%, 611/2.4\%, 338/1.3\%, and 211/0.84\%, respectively.


\begin{table}[!tbp]
\vspace{-0.58cm}
  \setlength\tabcolsep{1.3pt}
  \caption{DiaCorrect performance when pruning the training set.}
  \vspace{0.5mm}
  \label{tab:no_sas_aug}
  \centering
  \begin{tabular}{l|c|cccc|cccc}
    \hline
        \multirow{2}{*}{System} & \multirow{2}{*}{\textit{kept} \%} & \multicolumn{4}{c|}{CH Part2} & \multicolumn{4}{c}{DH3 eval} \\
        & & DER & Miss & FA & Conf. & DER & Miss & FA & Conf. \\
    \hline
    EEND-EDA & 100 &  8.62 & 3.40 & 4.53 & 0.69 & 19.58 & 3.94 & 14.62 & 1.02 \\
    \hline 
    \multirow{5}{*}{DiaCorrect} & 100 
        & 8.83 & 3.43 & 4.71 & 0.69 & 20.57  & 2.89 & 16.99 & 0.69 \\
        & 5.0 & 8.62 & 3.42 & 4.41 &  0.79 & 20.11 & 3.43 & 15.94 &    0.74 \\
        & 2.4 & 8.46 & 3.53 & 4.13 & 0.80 & 19.58 & 3.84 & 14.94 &   0.80 \\
        & 1.3 & 8.09 & 3.70 & 3.56 & 0.83 & 19.18 & 4.75 & 13.51 &    0.92 \\
        & 0.84 & 8.11 & 3.92 & 3.35 & 0.84 & 19.09 & 5.38 & 12.67 & 1.04 \\
    \hline
  \end{tabular}
   \vspace{-0.5cm}
\end{table}

Table \ref{tab:no_sas_aug} shows results of DiaCorrect when trained on each subset.
As we can see, without data pruning, the DiaCorrect performance is similar or even worse than the EEND-EDA baseline. 
When using smaller pruned sets for training, DiaCorrect outperforms the baseline.
The best performance is obtained with the two smallest 1.3\% and 0.84\% sets which attain comparable performance (yet smaller sets showed degradation).
While both sets were considered for further experiments, we observed worse results with the latter so we only use the 1.3\% set in the rest of this paper. By means of data pruning, we can decrease the initial DER on CH2 from 8.62\% to 8.09\% and DH3 eval from 19.58\% to 19.18\%.
To verify if such data selection could improve baseline performance as well, we also tried to further train the EEND-EDA baseline using the 1.3\% selected hard samples. However, with the same experimental setups as DiaCorrect, the CH2 DER performance changes from 8.62\% to 8.74\% and DH3 eval goes from 19.58\% to 19.62\%. 

\begin{figure}[!bp]
\vspace{-0.5cm}
  \centering
  \includegraphics[width=8.6cm]{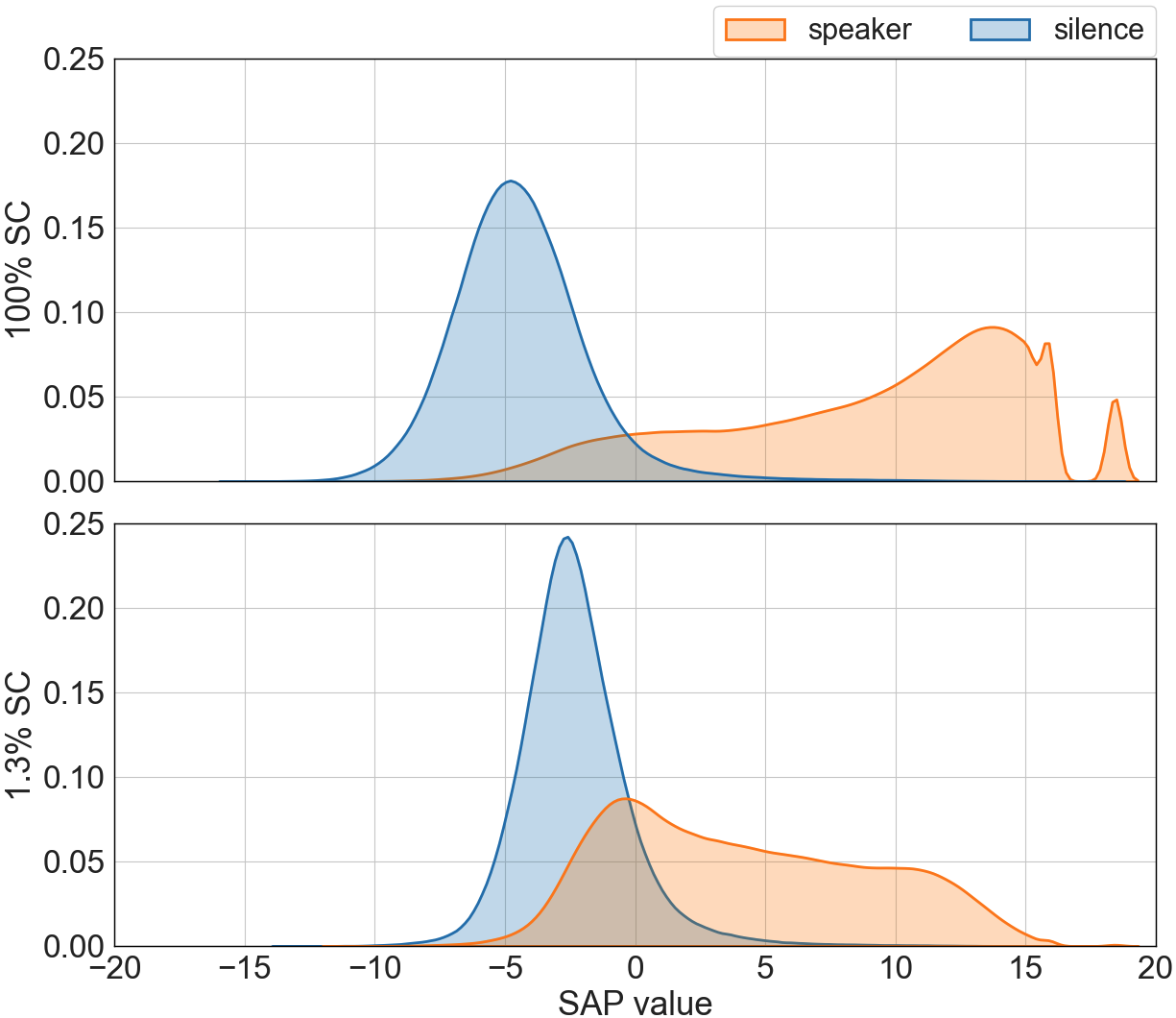}
  \caption{Initial SAP distributions of 100\% SC and 1.3\% SC.}
  \label{fig:sas_dist}
\end{figure}

To better understand the effect of the data pruning, Figure \ref{fig:sas_dist} shows the 
initial SAP distributions of different SC sets. For the 100\% SC set, we randomly select 1000 utterances. The distributions of speech and silence are obtained using the ground truth labels. In EEND-EDA, SAP values are passed through the sigmoid function to obtain output probabilities, thereby 0 corresponds to the 0.5 probability threshold used at inference time. With 0 serving as the decision boundary, a perfect diarization model would present completely disjunct distributions of silence and speech and, in such case, there would be no use for DiaCorrect. Consequently, correctly classified silence frames with very low values and speech frames with very high values do not provide useful information to train the error correction model. However, DiaCorrect can leverage the misclassified frames around 0 to improve performance.

\begin{figure}[!tbp]
\vspace{-0.5cm}
  \centering
  \includegraphics[width=8.6cm]{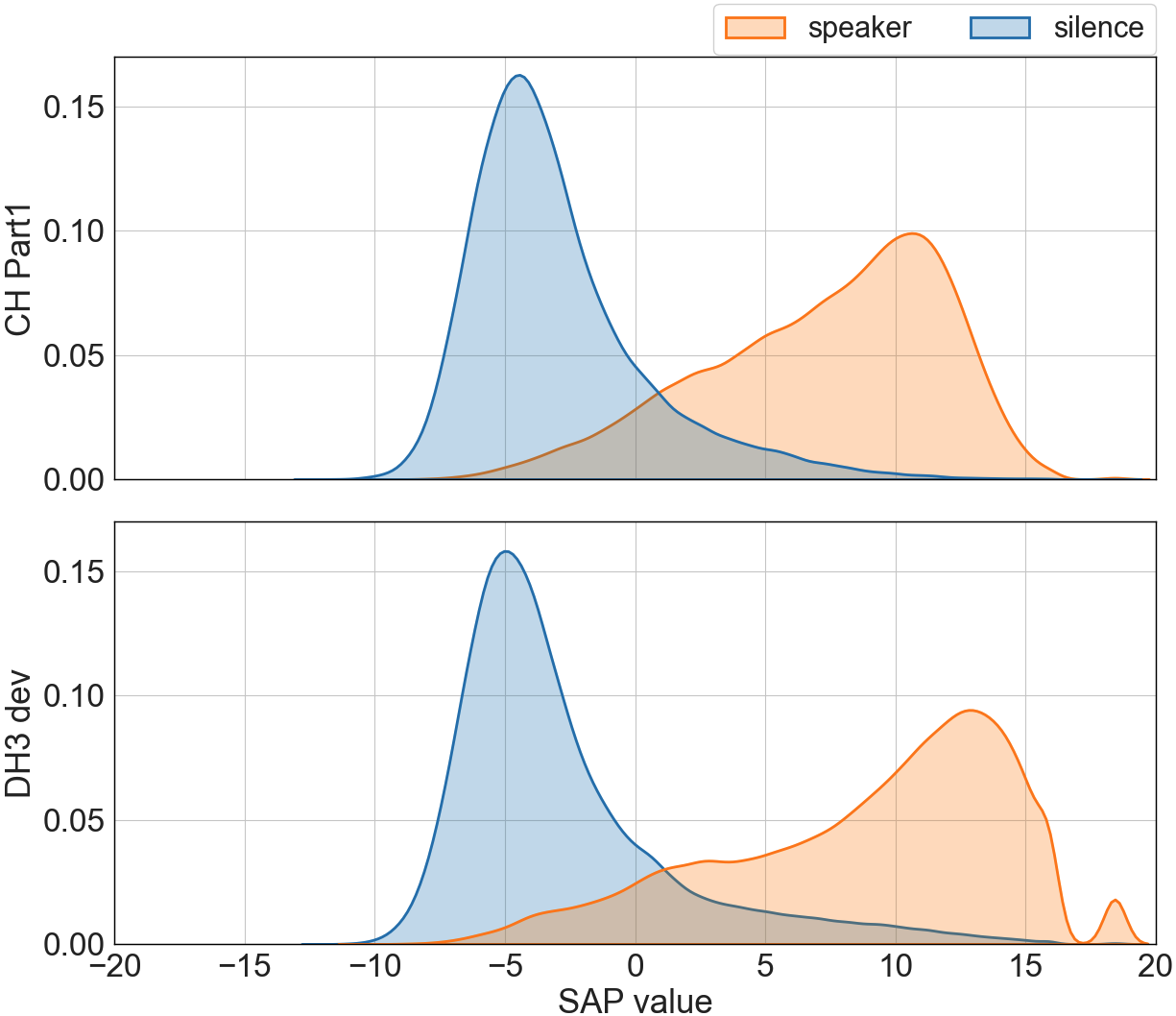}
  \caption{Initial SAP distribution of CH Part1 and DH3 dev.}
  \label{fig:ch1_dh3_dev_dist}
  \vspace{-0.2cm}
\end{figure}

\subsubsection{DiaCorrect with SAP calibration}
\label{sec:sap_calibration}
\begin{figure}[!tbp]
  \vspace{-0.5cm}
  \centering
  \includegraphics[width=8.8cm]{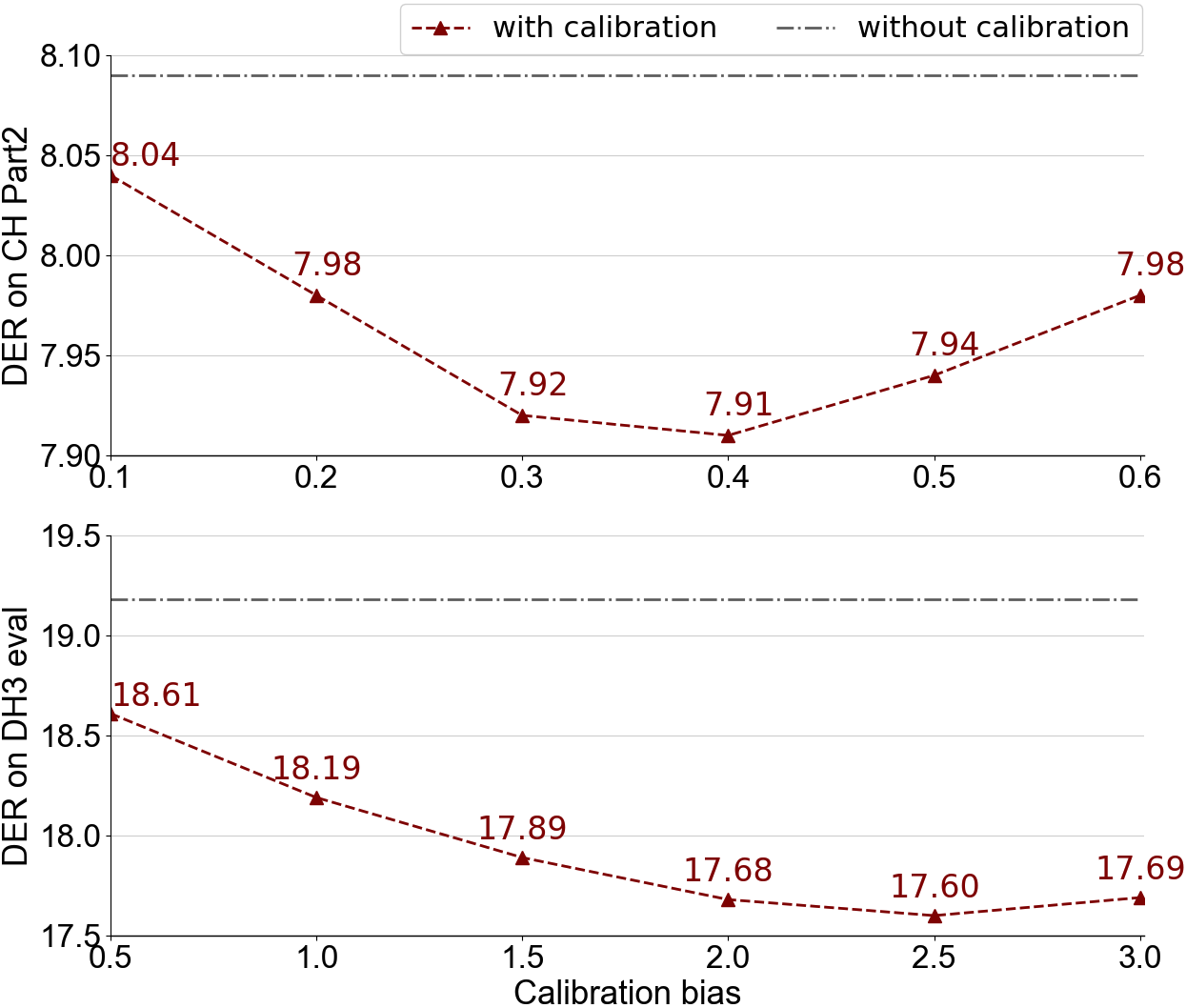}
  \caption{Effects of SAP calibration on CH Part2 and DH3 eval. 1.3\% pruned data is used for model training.}
  \label{fig:shift_line}
  \vspace{-0.25cm}
\end{figure}

\vspace{-0.1cm}
Although results in Table \ref{tab:no_sas_aug} indicate that DiaCorrect can bring improvements, the performance gain is limited, especially for DH3. To identify the potential reasons, in Figure \ref{fig:ch1_dh3_dev_dist}, we investigated the distributions for CH Part1 and DH3 dev. Given the distributions of silence and speech frames, the best decision threshold corresponds to the one that minimizes the combined misses and false alarms. While the theoretical threshold 0 performs well in the calibrated scenarios in Figure~\ref{fig:sas_dist}, it can be observed a different situation in Figure~\ref{fig:ch1_dh3_dev_dist}. To mitigate this mismatch, for the inference, we decided to calibrate the initial SAP distribution of real data. We investigated the effects of SAP calibration by simply subtracting some bias from its initial distributions. Figure \ref{fig:shift_line} shows the results of such calibration, which further improves the DiaCorrect performance. With only 30 hours of pruned simulated data, our system can reduce the initial DER on CH2 from 8.62\% to 7.91\% and DH3 from 19.58\% to 17.60\%, achieving similar performance on CH2 as the fine-tuned EEND-EDA.

\subsubsection{DiaCorrect with target domain data}
\vspace{-0.1cm}
Although manual labeling is expensive, it is not uncommon to have a small target-domain annotated corpus. Given the usual mismatch between synthetic training data and real recordings, fine-tuning to real data is essential to reach state-of-the-art performance with neural diarization methods \cite{fujita19_interspeech, fujita2019end, horiguchi20_interspeech, fujita2023neural}. In Table \ref{tab:ft_scratch}, we present the DiaCorrect performance when target data are used for training.
In this setup, no calibrations are needed, as training and inference are performed on data of the same domain.
As we can see, after fine-tuning, the initial DER can be further reduced. 
Especially on DH3 dev, the DER decreases from 17.60\% to 12.46\%. Compared to the fine-tuned results of EEND-EDA in Table \ref{tab:baseline}, our system achieves similar or even better results on both CH Part2 and DH3 eval subsets. This is promising for those scenarios where the diarization system is considered a black-box and the user does not have access to its internals. In this case, the fine-tuning of the diarization system is impossible, but as long as an inference interface and some real labeled data are provided, DiaCorrect can achieve competitive results.

In addition, we also train DiaCorrect from scratch using the target domain data (CH Part1 and DH3 dev).
Surprisingly, in this case, our system still achieves similar performance to the fine-tuned system. 
%
These results indicate that even when a few hours of target domain data are available, additional simulated data does not benefit the training of DiaCorrect.
Furthermore, note that the training from scratch of DiaCorrect requires only a few hours of labeled data, 3.19 hours for CH Part1 and 10.17 hours for DH3 dev, which is very friendly to computational resources.

\begin{table}[!htbp]
\vspace{-0.3cm}
  \setlength\tabcolsep{0.9pt}
  \caption{DiaCorrect performance when using labeled real data.}
  \vspace{0.5mm}
  \label{tab:ft_scratch}
  \centering
  \setlength{\tabcolsep}{1.8pt} 
  \begin{tabular}{l|cccc|cccc}
    \hline
        \multirow{2}{*}{System} & \multicolumn{4}{c|}{CH Part2} & \multicolumn{4}{c}{DH3 eval} \\
        & DER & Miss & FA & Conf. & DER & Miss & FA & Conf. \\
    \hline
      Train on SC & 7.91 & 4.11 & 2.96 & 0.84 & 17.60 & 7.39 & 9.15 & 1.06 \\
      \quad + FT on dev set & 7.85 & 4.74 & 2.30 & 0.81 & 12.46 & 7.10 & 4.71 & 0.65 \\
    Train on dev set & 7.98 & 4.67 & 2.76 & 0.55 & 12.47 & 7.11 & 4.64 & 0.72 \\
    \hline
  \end{tabular}
   \vspace{-0.4cm}
\end{table}

\vspace{-0.2cm}

\section{Conclusion}
\label{sec:conclusion}
\vspace{-0.1cm}

In this study, we propose 
DiaCorrect to refine the initial speaker diarization outputs of an existing system. 
We investigate data pruning and calibration to 
address over-fitting and data mismatch.
Experiments on 2-speaker CALLHOME and DIHARD-\uppercase\expandafter{\romannumeral3} CTS 
 demonstrate that 
our proposed DiaCorrect can significantly improve the initial diarization results. 
In particular, our system achieves promising performance when the diarization system is taken as a black-box, i.e. one cannot retrain the existing model.
Our future work will focus on exploring automatic (supervised and unsupervised) SAP calibration and generalizing DiaCorrect to more speakers. Furthermore, we plan to apply our method to other initial speaker diarization systems. 
\vspace{-0.1cm}
\section{Acknowledgements}
\label{sec:acknowledgements}
\vspace{-0.1cm}
The work was supported by Czech National Science Foundation (GACR) project NEUREM3 No.19-26934X, Czech Ministry of Interior project No.VJ01010108 "ROZKAZ", and Horizon 2020 Marie Sklodowska-Curie grant ESPERANTO, No.101007666. Computing on IT4I supercomputer was supported by the Czech Ministry of Education, Youth and Sports through the e-INFRA CZ (ID:90254).
\bibliographystyle{IEEEbib}
\bibliography{refs}

\begin{thebibliography}{10}

\bibitem{park2022review}
T.~J. Park, N.~Kanda, D.~Dimitriadis, K.~J. Han, S.~Watanabe, and S.~Narayanan,
\newblock ``A review of speaker diarization: Recent advances with deep
  learning,''
\newblock {\em Computer Speech \& Language}, vol. 72, pp. 101317, 2022.

\bibitem{sell2014speaker}
G.~Sell and D.~Garcia-Romero,
\newblock ``Speaker diarization with {PLDA} i-vector scoring and unsupervised
  calibration,''
\newblock in {\em Proc. SLT}. IEEE, 2014, pp. 413--417.

\bibitem{garcia2017speaker}
D.~Garcia-Romero, D.~Snyder, G.~Sell, D.~Povey, and A.~McCree,
\newblock ``Speaker diarization using deep neural network embeddings,''
\newblock in {\em Proc. ICASSP}. IEEE, 2017, pp. 4930--4934.

\bibitem{landini2022bayesian}
F.~Landini, J.~Profant, M.~Diez, and L.~Burget,
\newblock ``Bayesian {HMM} clustering of x-vector sequences ({VBx}) in speaker
  diarization: theory, implementation and analysis on standard tasks,''
\newblock {\em Computer Speech \& Language}, vol. 71, pp. 101254, 2022.

\bibitem{fujita19_interspeech}
Y.~Fujita, N.~Kanda, S.~Horiguchi, K.~Nagamatsu, and S.~Watanabe,
\newblock ``End-to-end neural speaker diarization with permutation-free
  objectives,''
\newblock in {\em Proc. Interspeech}, 2019, pp. 4300--4304.

\bibitem{fujita2019end}
Y.~Fujita, N.~Kanda, S.~Horiguchi, Y.~Xue, K.~Nagamatsu, and S.~Watanabe,
\newblock ``End-to-end neural speaker diarization with self-attention,''
\newblock in {\em Proc. ASRU}, 2019, pp. 296--303.

\bibitem{horiguchi20_interspeech}
S.~Horiguchi, Y.~Fujita, S.~Watanabe, Y.~Xue, and K.~Nagamatsu,
\newblock ``End-to-end speaker diarization for an unknown number of speakers
  with encoder-decoder based attractors,''
\newblock in {\em Proc. Interspeech}, 2020, pp. 269--273.

\bibitem{fujita2023neural}
Y.~Fujita, T.~Komatsu, R.~Scheibler, Y.~Kida, and T.~Ogawa,
\newblock ``Neural diarization with non-autoregressive intermediate
  attractors,''
\newblock in {\em Proc. ICASSP}. IEEE, 2023, pp. 1--5.

\bibitem{yamashita22_odyssey}
Natsuo Yamashita, Shota Horiguchi, and Takeshi Homma,
\newblock ``{Improving the Naturalness of Simulated Conversations for
  End-to-End Neural Diarization},''
\newblock in {\em Proc. The Speaker and Language Recognition Workshop (Odyssey
  2022)}, 2022, pp. 133--140.

\bibitem{landini2022simulated}
F.~Landini, A.~Lozano-Diez, M.~Diez, and L.~Burget,
\newblock ``From simulated mixtures to simulated conversations as training data
  for end-to-end neural diarization,''
\newblock in {\em Proc Interspeech}, 2022, pp. 5095--5099.

\bibitem{landini2023multi}
F.~Landini, M.~Diez, A.~Lozano-Diez, and L.~Burget,
\newblock ``Multi-speaker and wide-band simulated conversations as training
  data for end-to-end neural diarization,''
\newblock in {\em Proc. ICASSP}. IEEE, 2023, pp. 1--5.

\bibitem{errattahi2018automatic}
R.~Errattahi, A.~El~Hannani, and H.~Ouahmane,
\newblock ``Automatic speech recognition errors detection and correction: A
  review,''
\newblock {\em Procedia Computer Science}, vol. 128, pp. 32--37, 2018.

\bibitem{leng2021fastcorrect}
Y.~Leng, X.~Tan, L.~Zhu, J.~Xu, R.~Luo, L.~Liu, T.~Qin, X.~Li, E.~Lin, and
  T.-Y. Liu,
\newblock ``Fastcorrect: Fast error correction with edit alignment for
  automatic speech recognition,''
\newblock {\em Advances in Neural Information Processing Systems}, vol. 34, pp.
  21708--21719, 2021.

\bibitem{guo2019spelling}
J.~Guo, T.~N. Sainath, and R.~J. Weiss,
\newblock ``A spelling correction model for end-to-end speech recognition,''
\newblock in {\em Proc. ICASSP}. IEEE, 2019, pp. 5651--5655.

\bibitem{zhang2019investigation}
S.~Zhang, M.~Lei, and Z.~Yan,
\newblock ``Investigation of transformer based spelling correction model for
  {CTC}-based end-to-end {Mandarin} speech recognition.,''
\newblock in {\em Proc. Interspeech}, 2019, pp. 2180--2184.

\bibitem{zhang2021end}
S.~Zhang, J.~Yi, Z.~Tian, Y.~Bai, J.~Tao, X.~Liu, and Z.~Wen,
\newblock ``End-to-end spelling correction conditioned on acoustic feature for
  code-switching speech recognition.,''
\newblock in {\em Interspeech}, 2021, pp. 266--270.

\bibitem{hrinchuk2020correction}
Oleksii Hrinchuk, Mariya Popova, and Boris Ginsburg,
\newblock ``Correction of automatic speech recognition with transformer
  sequence-to-sequence model,''
\newblock in {\em Proc. ICASSP}. IEEE, 2020, pp. 7074--7078.

\bibitem{horiguchi2021end}
S.~Horiguchi, P.~Garcia, Y.~Fujita, S.~Watanabe, and K.~Nagamatsu,
\newblock ``End-to-end speaker diarization as post-processing,''
\newblock in {\em Proc. ICASSP}. IEEE, 2021, pp. 7188--7192.

\bibitem{paturi2023lexical}
R.~Paturi, S.~Srinivasan, and X.~Li,
\newblock ``Lexical speaker error correction: Leveraging language models for
  speaker diarization error correction,''
\newblock in {\em Proc. Interspeech}, 2023, pp. 3567--3571.

\bibitem{callhome}
``2000 {NIST} speaker recognition evaluation,'' \url{https://catalog.ldc.
  upenn.edu/LDC2001S97}.

\bibitem{ryant21_interspeech}
R.~Neville, S.~Prachi, K.~Venkat, V.~Rajat, C.~Kenneth, C.~Christopher, D.~Jun,
  G.~Sriram, and L.~Mark,
\newblock ``The third {DIHARD} diarization challenge,''
\newblock in {\em Proc. Interspeech}, 2021, pp. 3570--3574.

\bibitem{kolbaek2017multitalker}
M.~Kolb{\ae}k, D.~Yu, Z.-H. Tan, and J.~Jensen,
\newblock ``Multitalker speech separation with utterance-level permutation
  invariant training of deep recurrent neural networks,''
\newblock {\em IEEE/ACM Transactions on Audio, Speech, and Language
  Processing}, vol. 25, no. 10, pp. 1901--1913, 2017.

\bibitem{luo2019conv}
Y.~Luo and N.~Mesgarani,
\newblock ``Conv-tasnet: Surpassing ideal time--frequency magnitude masking for
  speech separation,''
\newblock {\em IEEE/ACM transactions on audio, speech, and language
  processing}, vol. 27, no. 8, pp. 1256--1266, 2019.

\bibitem{chollet2017xception}
F.~Chollet,
\newblock ``Xception: Deep learning with depthwise separable convolutions,''
\newblock in {\em Proc. CVPR}, 2017, pp. 1251--1258.

\bibitem{he2015delving}
K.~He, X.~Zhang, S.~Ren, and J.~Sun,
\newblock ``Delving deep into rectifiers: Surpassing human-level performance on
  imagenet classification,''
\newblock in {\em Proc. ICCV}, 2015, pp. 1026--1034.

\bibitem{ba2016layer}
J.~L. Ba, J.~R. Kiros, and G.~E. Hinton,
\newblock ``Layer normalization,''
\newblock {\em arXiv preprint arXiv:1607.06450}, 2016.

\bibitem{he2016deep}
K.~He, X.~Zhang, S.~Ren, and J.~Sun,
\newblock ``Deep residual learning for image recognition,''
\newblock in {\em Proc. CVPR}, 2016, pp. 770--778.

\bibitem{vaswani2017attention}
A.~Vaswani, N.~Shazeer, N.~Parmar, J.~Uszkoreit, L.~Jones, A.~N. Gomez,
  {\L}.~Kaiser, and I.~Polosukhin,
\newblock ``Attention is all you need,''
\newblock {\em Advances in neural information processing systems}, vol. 30,
  2017.

\bibitem{kingma2014adam}
D.~P. Kingma and J.~Ba,
\newblock ``Adam: A method for stochastic optimization,''
\newblock {\em arXiv preprint arXiv:1412.6980}, 2014.

\bibitem{fiscus2006rich}
J.~G. Fiscus, J.~Ajot, M.~Michel, and J.~S. Garofolo,
\newblock ``The rich transcription 2006 spring meeting recognition
  evaluation,''
\newblock in {\em International Workshop on Machine Learning for Multimodal
  Interaction}. Springer, 2006, pp. 309--322.

\end{thebibliography}
\end{document}